\documentclass[
  reprint,
  superscriptaddress,
  amsmath,amssymb,
  aps,
]{revtex4-2}

\usepackage{graphicx}
\usepackage{dcolumn}
\usepackage{bm}

\usepackage{booktabs}
\usepackage{graphicx}

\usepackage[utf8]{inputenc}
\usepackage{graphicx}
\usepackage{amsmath,amssymb}

\usepackage{xcolor}
\usepackage{hyperref}

\usepackage[utf8]{inputenc}
\usepackage{graphicx} 
\usepackage{amsmath}  
\usepackage{amssymb}  

\usepackage{geometry} 
\usepackage{array}    
\usepackage{multirow} 
\usepackage{placeins} 

\usepackage{booktabs} 
\usepackage{threeparttable} 
\usepackage{xcolor} 
\usepackage{hyperref} 
\usepackage{makecell} 
\usepackage{siunitx}  
\usepackage{xcolor}

\begin{document}

\title{Fano-Like Resonances in Coupled Sagnac Interferometers Formed by a Self-Coupled Waveguide}

\author{Hamed Arianfard}
\affiliation{Quantum Photonics Laboratory and Centre for Quantum Computation and Communication Technology, RMIT University, Melbourne, VIC 3000, Australia}
\author{Tim Weiss}
\affiliation{Quantum Photonics Laboratory and Centre for Quantum Computation and Communication Technology, RMIT University, Melbourne, VIC 3000, Australia}
\author{Yang Yang}
\affiliation{Quantum Photonics Laboratory and Centre for Quantum Computation and Communication Technology, RMIT University, Melbourne, VIC 3000, Australia}
\author{Joshua Bader}
\affiliation{School of Engineering, RMIT University, Melbourne, 3000, VIC, Australia}
\author{Stefania Castelletto}
\affiliation{School of Engineering, RMIT University, Melbourne, 3000, VIC, Australia}
\author{Alberto Peruzzo}
\email[]{alberto.peruzzo@gmail.com}

\thanks{\\This work has been submitted to the IEEE for possible publication. 
  Copyright may be transferred without notice, after which this version may no longer be accessible.}
  
\affiliation{Quantum Photonics Laboratory and Centre for Quantum Computation and Communication Technology, RMIT University, Melbourne, VIC 3000, Australia}
\affiliation{Quandela, Massy, France}

\begin{abstract}
We demonstrate Fano-like resonances in silicon-on-insulator (SOI) nanowire-based coupled Sagnac interferometers (SIs) formed by a self-coupled waveguide. By adjusting the reflectivity of the two SIs and coupling strength between them, we tailor coherent mode interference to achieve high-performance optical analogues of Fano resonance. The device is theoretically analyzed and experimentally fabricated on a SOI platform. Theoretical analysis predicts periodic Fano-like resonances with a high extinction ratio and a steep slope rate, arising from strong coherent optical mode interference within a compact resonator comprising two SIs and a connected feedback waveguide. Experimental results align with the theoretical model, validating the expected resonance behavior and confirming the effectiveness of the design. These findings underscore the potential of compact coupled SIs for generating Fano-like resonances, enabling broader applications in integrated photonics.
\end{abstract}
\maketitle

\section{Introduction}

Integrated photonic resonators (IPRs) are fundamental to advancing modern photonic systems, offering critical functionalities for applications ranging from telecommunications to quantum technologies \cite{chrostowski2015silicon, rabus2007integrated, bogaerts2012silicon}. Among various resonator designs, integrated Sagnac interferometers (SIs) stand out due to their unique advantages. These devices support bidirectional light propagation and mutual coupling between counter-propagating paths, facilitating complex coherent mode interference and precise spectral control \cite{arianfard2023sagnac}. Unlike traveling-wave resonators, such as ring resonators (RRs), SIs use standing-wave resonators to achieve a compact footprint without sacrificing robust resonant interactions \cite{arianfard2023sagnac}. This enables increased quality (Q) factors and narrower free spectral ranges (FSRs). While a narrower FSR may limit channel count in dense wavelength-division multiplexing (DWDM) systems, it is advantageous for applications requiring dense spectral resolution, noise suppression, and high integration density. Moreover, devices can be designed to meet the standard DWDM channel spacings specified by the ITU-T G.694.1 spectral grid \cite{barea2013silicon,ITU-G694.1}. These features position SI-based devices as versatile components for both classical and quantum photonics, where key performance metrics like spectral resolution and integration density are crucial \cite{shekhar2024roadmapping, wang2020integrated}.

Fano resonance, first introduced by Ugo Fano in 1935, describes the asymmetric spectral line shapes resulting from the interference between a discrete quantum state and a continuum of states \cite{fano1935sullo, fano1961effects}. While initially studied in the context of atomic physics, the concept has since been extended to various fields, including condensed matter physics, nanophotonics, and quantum technologies, demonstrating its universal relevance in understanding interference effects \cite{miroshnichenko2010fano, limonov2017fano, limonov2021fano}. Fano resonances are of particular interest for their asymmetric line shapes, which, under favourable conditions and careful device implementation, can yield ultra-narrow linewidths and high extinction ratios (ERs), although these properties are not universal. These characteristics, surpassing the capabilities of conventional Lorentzian lineshapes, have enabled the realization of various proof-of-concept devices, including optical modulators \cite{yu2015ultrafast, piao2012control}, lasers \cite{yu2017demonstration, rasmussen2018modes}, spasers \cite{zheng2015low, huo2014spaser}, sensors \cite{meng2018control, yi2010highly}, switches \cite{yu2014fano, stern2014fano}, and many \cite{arruda2018fano, doeleman2020observation, vabishchevich2018enhanced, limonov2017fano, limonov2021fano}.

Recent innovations have demonstrated that various schemes, including photonic crystals \cite{bekele2019plane}, nanoclusters \cite{king2015fano, zhang2014coherent}, gratings \cite{hua2019tunable, sounas2018fundamental}, metamaterials \cite{papasimakis2009metamaterial, lim2018ultrafast}, and metasurfaces \cite{cui2018multiple}, can exhibit Fano resonances. Integrating Fano resonances into IPRs offers several competitive benefits, including compact footprints, high scalability, robust stability, and the potential for mass production \cite{zhang2016optically, li2017actively, zhao2019independently}. Our earlier work theoretically explored photonic filters based on various SI-based configurations \cite{arianfard2020advanced, arianfard2021three, arianfard2021spectral, arianfard2023optical}. Here, we experimentally validate these theoretical insights using a compact resonator composed of two coupled SIs and a feedback waveguide, based on self-coupled waveguide concepts \cite{lai2016compact}, fabricated on a silicon-on-insulator (SOI) platform. Specifically, we aim to realize Fano-like resonances, where theoretical investigations predict that a maximum ER of \(\sim 63\) dB and a maximum slope rate (SR) of \(\sim 948\) dB/nm, along with a low insertion loss (IL) of \(\sim 0.57\) dB, can be achieved by adjusting the reflectivity of the two SIs and the coupling strength between them. The measured results exhibit the spectral response characteristics of the device predicted by the theoretical model. However, discrepancies between the measured and theoretical ERs and SRs primarily arise from resolution limitations in the laser scanning measurement system. A vector network analyzer-based experimental setup would provide a more accurate method for measuring Fano-like resonances \cite{liu2019tunable, cheng2022achieving}. High ER and SR Fano resonances in integrated photonics are critical for advanced applications. A high ER is essential for optical switching, noise suppression, and precise filtering in optical communication systems, while a steep SR offers significant advantages for high-sensitivity applications such as biosensing, chemical sensing, and nonlinear photonic processes \cite{luk2010fano, yi2010highly, limonov2017fano, miroshnichenko2010fano, stern2014fano}. The combination of a high ER, high SR, along with a low insertion loss (IL) of $\sim 0.57 \, \text{dB}$ (obtained from theoretical calculations), and a compact, simple structure, compared to various Fano-resonance-based schemes on the SOI platform \cite{yu2013fano, hu2013tunable, zhang2016optically, zhao2016tunable, wang2016fano, zheng2017compact, wang2017slope, li2017actively, zhao2019independently, liu2019tunable}, positions this device as a promising candidate for Fano-resonance applications high-performance sensing, optical modulation, and photonic signal processing. In addition, schemes that combine RRs with Fabry-Pérot (FP) cavities \cite{zheng2017compact} or Mach–Zehnder interferometers (MZIs) \cite{liu2019tunable} not only impose stringent requirements on the alignment of resonant wavelengths across the individual subcomponents but also result in a bulkier device footprint. Furthermore, maintaining the desired spectral response is challenging due to unequal wavelength drifts among the subcomponents induced by the thermo-optic effect. Unlike conventional hybrid structures, the proposed self-coupled waveguide–based architecture achieves a compact, high-performance design that eliminates alignment challenges and enhances device robustness.

\section{Design and principle}

As shown schematically in Figure 1(a), the interaction between a discrete state $\lvert d \rangle$ and a continuum $\lvert c \rangle$ gives rise to a Fano resonance, characterized by an asymmetric line shape. 
\begin{figure*}[!htb]
    \centering
    \includegraphics[width=\textwidth]{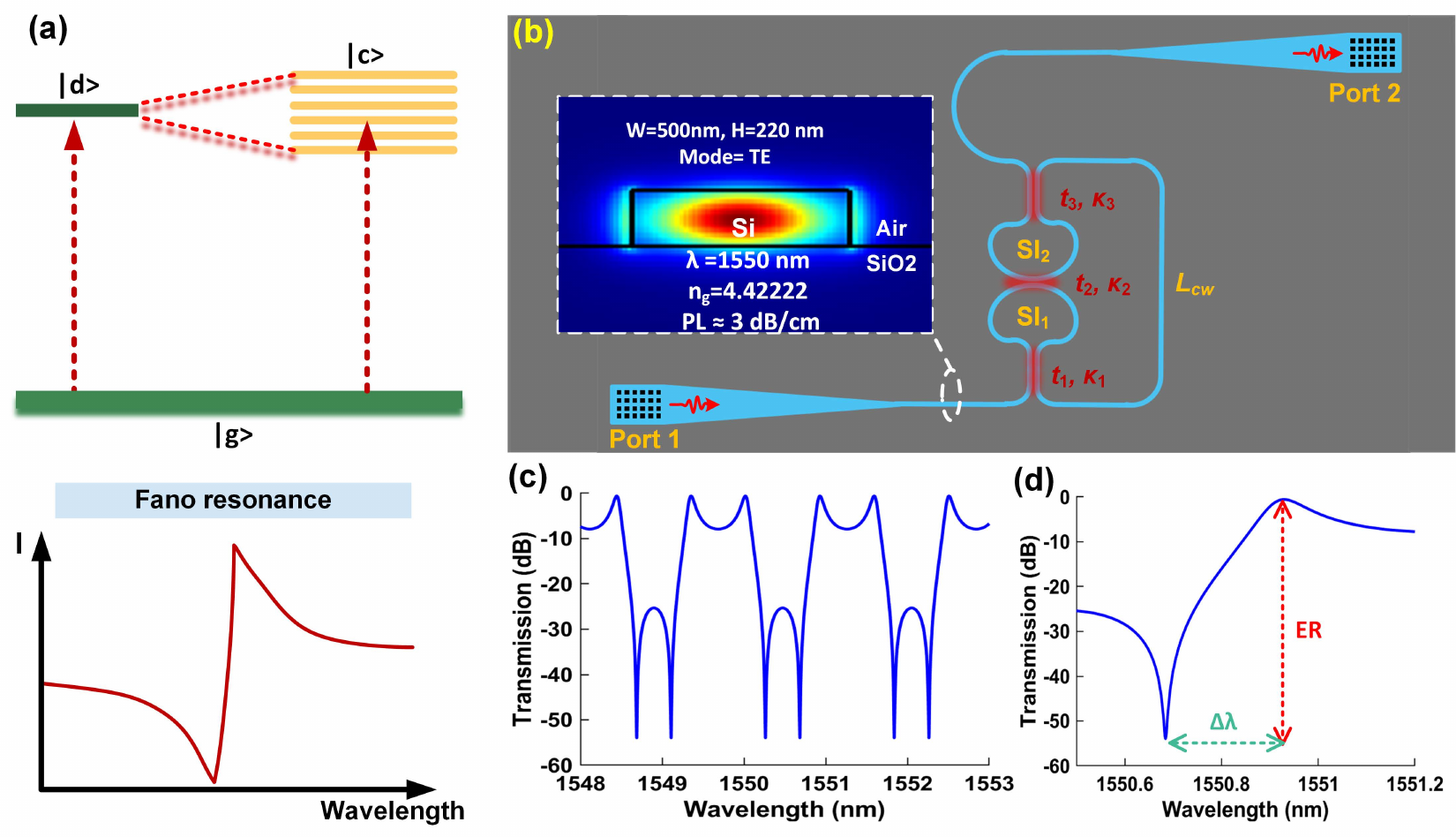}
    \caption{(a) Schematic illustration of Fano resonance. \(|g\rangle\): ground state. \(|d\rangle\): a discrete state. \(|c\rangle\): a continuum of states. (b) Schematic configuration of the device consisting of coupled SIs formed by a self-coupled wire waveguide. The definitions of \(t_i \; (i = 1{-}3), k_i \; (i = 1{-}3), \text{SI}_i \; (i = 1, 2), L_{cw}\) are provided in Table~\ref{tab:device_parameters}. The inset shows the cross-section of the simulated fundamental TE mode at $\lambda = 1550~\text{nm}$. W: width. H: height. \(n_g\): group index. PL: propagation loss.  (c) Transmission spectrum of the device at Port 2, where $t_1 = t_3 = 0.88$, $t_2 = 0.98$, $L_{s_1} = L_{s_2} = 115~\mu$m, and $L_{\text{cw}} = 230~\mu$m. (d) Zoom-in view of (c) in the wavelength range of 1550.5~nm--1551.2~nm.  $\Delta\lambda$: wavelength difference between the resonance peak and notch. ER: extinction ratio. }
    \label{fig:design}
\end{figure*}
A discrete energy state (e.g., an atom, a quantum dot, or a localized photonic mode) interacting with a broad continuum of energy states (e.g., free-electron states, propagating photonic waveguide modes, or delocalized plasmonic bands) is the fundamental mechanism underlying Fano resonance. This interaction gives rise to interference between the narrow spectral response of the discrete state and the broad, continuous background of the continuum. As a result, the system exhibits an asymmetric spectral line shape, which is a hallmark of Fano resonance.

Figure 1(b) presents a schematic of the proposed compact resonator, which incorporates two SIs and a feedback waveguide. The structural parameters of the device are defined in Table~\ref{tab:device_parameters}. The spectral response of the device is calculated using the scattering matrix method~\cite{arianfard2023sagnac,arianfard2020advanced}. For the calculation, we assume a transverse electric (TE) mode group index of \( n_g = 4.42222 \) (obtained via a frequency-domain eigenmode solver for a single-mode silicon photonic nanowire waveguide with a cross-section of 500 nm × 220 nm, as shown in the inset of Figure 1(b)) and a propagation loss of \( \alpha = 70 \, \text{m}^{-1} \) (equivalent to \(\sim 3 \, \text{dB/cm}\)), in line with previously fabricated SOI devices \cite{wu2014compact, dumon2004low}. To calculate the spectral response of the device shown in Figure 1(b) using the scattering matrix method, the device was segmented into individual directional couplers and connecting waveguides. Scattering matrix equations were derived to describe the relationships between the input and output optical fields at these components. The system was initialized with an input optical field of 1 at Port 1, and the output spectral transfer function at Port 2 was determined by solving the resulting set of linear equations. Further details on this method are available in Refs.~\cite{arianfard2023sagnac,arianfard2020advanced}. It should be noted that the directional couplers are modeled using the universal unitary scattering matrix under the assumption of lossless coupling~\cite{yariv2000universal, yariv2002critical}. In practical devices, however, coupling loss is present and leads to a slight increase in the overall insertion loss. Moreover, the coupling strength of silicon directional couplers depends only weakly on wavelength, as detailed in Ref.~\cite{arianfard2023sagnac}, across the telecom C-band (1530--1565~nm). Over the narrow free spectral range of our device, this variation is negligible. In contrast, for broadband devices, the coupling strength becomes wavelength-dependent due to material and structural dispersion. Although narrowing the waveguide gap can mitigate this dispersion-induced effect, fabrication constraints limit the minimum achievable gap width. Therefore, various alternative coupler designs have been proposed to minimize wavelength dependence in integrated optical circuits~\cite{arianfard2023sagnac}.

\begin{table*}[t]
\centering
\caption{Definitions of device structural parameters.}
\resizebox{\textwidth}{!}{%
\begin{tabular}{@{}p{5cm}p{5cm}p{5cm}p{5cm}@{}}
\toprule
\textbf{Waveguides} & \textbf{Physical length} & \textbf{Transmission factor\textsuperscript{a}} & \textbf{Phase shift\textsuperscript{b}} \\
\midrule
Connecting waveguide between SI$_1$ and SI$_2$ (feedback waveguide)
& $L_{\mathrm{cw}}$
& $a_{\mathrm{cw}} = \exp(-\alpha L_{\mathrm{cw}} / 2)$
& $\phi_{\mathrm{cw}} = 2\pi n_g L_{\mathrm{cw}} / \lambda$ \\[3pt]

Sagnac loop in SI$_i$ ($i = 1, 2$)
& $L_{\mathrm{si}}$
& $a_{\mathrm{si}} = \exp(-\alpha L_{\mathrm{si}} / 2)$
& $\phi_{\mathrm{si}} = 2\pi n_g L_{\mathrm{si}} / \lambda$ \\[3pt]
\midrule
\textbf{Directional couplers} & & \textbf{Field transmission coefficient\textsuperscript{c}} & \textbf{Field cross-coupling coefficient\textsuperscript{c}} \\[3pt]

Coupler in SIs & & $t_i$ ($i = 1, 3$) & $\kappa_i$ ($i = 1, 3$)\\[3pt]
Coupler between SI$_1$ and SI$_2$ & & $t_2$ & $\kappa_2$ \\
\bottomrule
\end{tabular}%
}
\vspace{0.75em}

\parbox{\textwidth}{%
\footnotesize
\textsuperscript{a}$a_{\mathrm{cw}} = \exp(-\alpha L_{\mathrm{cw}} / 2)$ and $a_{\mathrm{si}} = \exp(-\alpha L_{\mathrm{si}} / 2)$, where $\alpha$ is the power propagation loss factor.\\
\textsuperscript{b}$\phi_{\mathrm{cw}} = 2\pi n_g L_{\mathrm{cw}} / \lambda$ and $\phi_{\mathrm{si}} = 2\pi n_g L_{\mathrm{si}} / \lambda$, where $n_g$ is the group index and $\lambda$ is the wavelength.\\
\textsuperscript{c}$t_i^2 + \kappa_i^2 = 1$ and $t_2^2 + \kappa_2^2 = 1$ are assumed for lossless directional couplers. SI: Sagnac interferometer.
}
\label{tab:device_parameters}
\end{table*}

We tailor coherent mode interference
within the resonator shown in Figure 1(b), to achieve optical analogues of Fano resonance. The flexibility in tuning the reflectivity of the SIs (i.e., \(t_1 , t_3\)), the coupling strength between the two SIs (i.e., \(t_2\)), as well as the lengths of the SIs (i.e., \(L_{S1}, L_{S2}\)) and the connecting waveguide (i.e., \(L_{cw}\)), forms the foundation for tailoring the spectral response of the device, which leads to the generation of Fano-like resonances. The resonator is equivalent to two cascaded SIs, forming a FP cavity (which acts as an infinite-impulse-response (IIR) filter) when \( t_2 = 1 \), and a single SI (which functions as a finite-impulse-response (FIR) filter) when \( t_1 = t_3 = 1 \). When \( t_i \) (\( i = 1\text{--}3 \)) \( \neq 1 \), the device operates as a hybrid filter, combining both IIR and FIR elements. This configuration enables flexible coherent mode interference, allowing for a diverse range of spectral responses. In this work, we specifically target the realization of Fano-like resonances.

The power transmission spectrum at Port 2 with light input from Port 1 are illustrated in Figure 1(c). In Figure 1(c), the structural parameters of the device are defined as $t_{1} = t_{3} = 0.88$, $t_{2} = 0.98$, $L_{s1} = L_{s2} = 115$ \textmu m, and $L_{cw} = 230$ \textmu m. Clearly, the transmission spectrum at Port 2 shows periodic Fano-like resonances with an asymmetric resonant lineshape in each period. The consistent uniformity of the filter shape across multiple periods or channels is highly beneficial for  wavelength division multiplexing systems. A zoomed-in view of Figure 1(c) is presented in Figure 1(d). The Fano-like resonance shown in Figure 1(d) exhibit a high ER of $\sim 53.41 \, \text{dB}$, a SR—defined as the ratio of the ER to the wavelength separation between the resonance peak and the notch—of $\sim 218 \, \text{dB/nm}$, and a insertion loss (IL) of $\sim 0.57 \, \text{dB}$. The SR definition is adopted for consistency with prior studies and provides a reliable metric for comparing the sharpness and sensitivity of Fano resonances across different device designs. Although the Fano lineshape is inherently nonlinear, the local slope near the inflection point can be used for more precise evaluation in specific sensing or modulation application. Notably, an IL of of $\sim 0.57 \, \text{dB}$, is outstanding among reported Fano-resonance devices on the SOI platform \cite{li2017actively, zhao2019independently}, making this device highly appealing for practical optical communication systems.
\begin{figure*}[!htb]
    \centering
    \includegraphics[width=\textwidth]
    {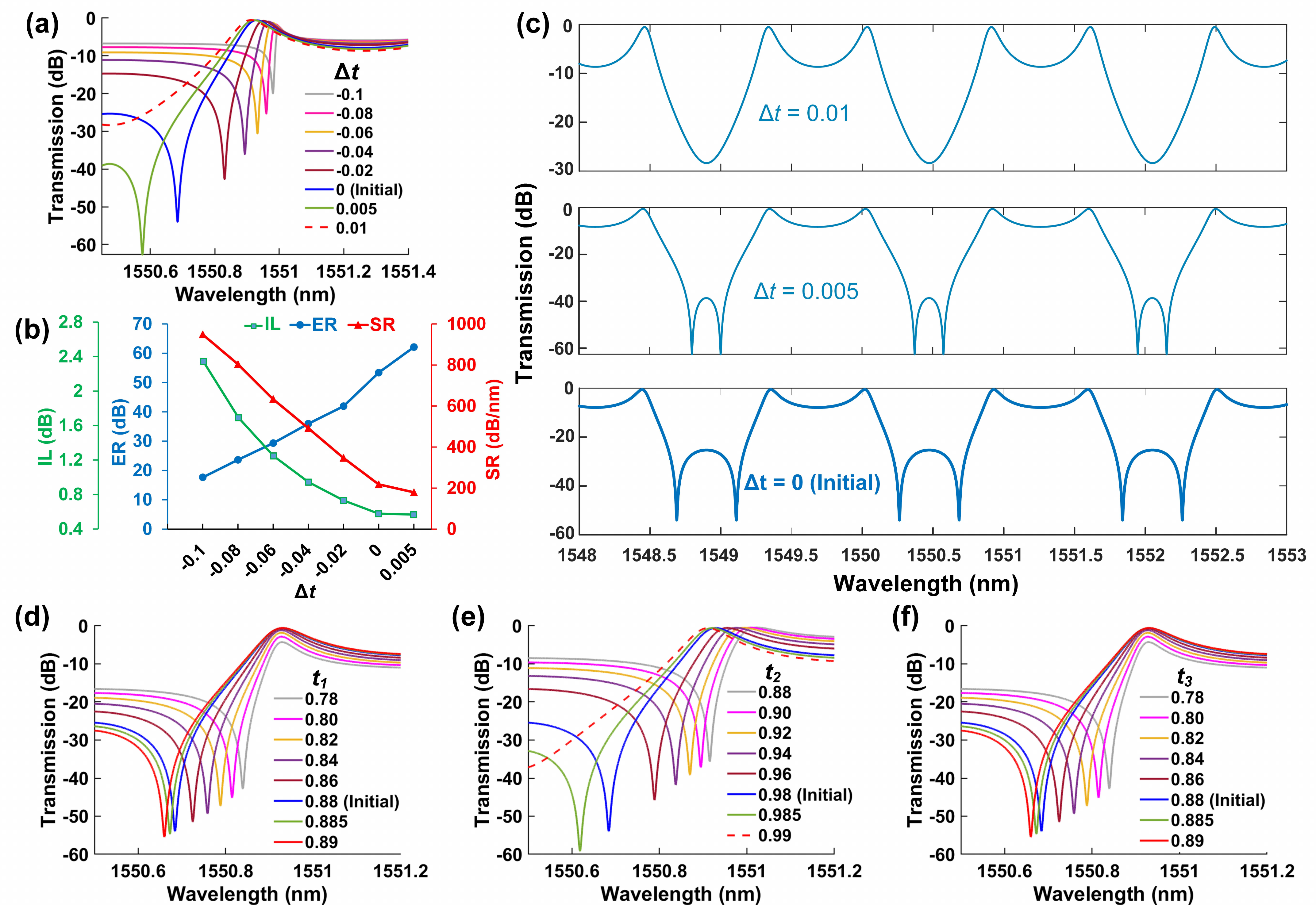} 
    \caption{Influence of the variation in coupling strength of directional couplers on the device spectral response. (a) Power transmission spectra for various \(\Delta t\) at Port 2. (b) Calculated IL, ER, and SR as functions of \(\Delta t\) for the Fano-like resonances in (a). (c) The transmission spectra in Figure 2(a) over a wider wavelength range of 1548 nm–1553 nm for three values of \(\Delta t = 0\), \(0.005\), and \(0.01\). In (a)--(c), \(\Delta t\) denotes a common offset applied to the field transmission coefficient of each of the three directional couplers (\(t_i \rightarrow t_i + \Delta t\) for \(i = 1\text{--}3\)). (d)–(f) Power transmission spectra for different \(t_i\) (\(i = 1\)–\(3\)), respectively. In (a)--(f), the “initial value” curves correspond to the baseline design with \(t_1 = t_3 = 0.88\), \(t_2 = 0.98\), \(L_{s1} = L_{s2} = 115\,\mu\mathrm{m}\), and \(L_{cw} = 230\,\mu\mathrm{m}\) (as in Fig.~1(c)). All other structural parameters are identical to those in Fig.~1(c), except for the ones being varied.}
    \label{fig:Simu}
\end{figure*}

In Figure 2, we investigate the impact of coupling-strength variations in the directional couplers on the device spectral response. Figure~2(a) compares the power transmission spectra for various values of $\Delta t$. To model uniform variations in directional-coupler strength (e.g., caused by bending sections or systematic fabrication offsets), we apply a common offset $\Delta t$ to the field transmission coefficients of all three couplers ($t_i \rightarrow t_i + \Delta t$ for $i = 1$--$3$). The baseline case ($\Delta t = 0$) uses $t_1 = t_3 = 0.88$, $t_2 = 0.98$, $L_{s1} = L_{s2} = 115~\mu\mathrm{m}$, and \(L_{cw} = 230\,\mu\mathrm{m}\), identical to the design in Fig.~1(c). The calculated IL, ER, and SR as functions of $\Delta t$ for the Fano-like resonances in Fig.~2(a) are shown in Fig.~2(b). Clearly, the ER increases with $\Delta t$, while the SR decreases, highlighting the trade-off between them. The IL shows a gradual decrease as $\Delta t$ increases. A maximum value of \(\sim 63\, \text{dB}\) for the ER is achieved at \(\Delta t = 0.005\), with a corresponding SR of \(\sim 179.2\, \text{dB/nm}\) and an IL of \(\sim 0.56\, \text{dB}\), while a maximum value of \(\sim 948\, \text{dB/nm}\) for the SR is achieved at \(\Delta t = -0.1\), with a corresponding ER of \(\sim 11.2\, \text{dB}\) and IL of \(\sim 2.34\, \text{dB}\). With its high ER, high SR, low IL, and simple structure, this device stands out as a promising candidate for Fano-resonance applications on the SOI platform \cite{yu2013fano, hu2013tunable, zhang2016optically, zhao2016tunable, wang2016fano, zheng2017compact, wang2017slope, li2017actively, zhao2019independently, liu2019tunable}.
Figure 2(c) shows the power transmission spectrum for three different values of \(\Delta t\). As \(\Delta t\) increases, the spectral range between the two split resonances in the stop band of the transmission spectrum gradually decreases and merges into a single resonance (when \(\Delta t = 0.01\)). Consequently, the Fano asymmetric spectral lineshape evolves into a horn-like spectral lineshape with a decreased notch depth. This illustrates how transitions between Fano and horn-like resonance lineshapes can be achieved by simply adjusting \(\Delta t\). Figures~2(d)--(f) show the effect of individually varying each directional coupler transmission coefficient ($t_1$, $t_2$, and $t_3$) while keeping the other parameters fixed, as in Fig.~1(c). As $t_1$ or $t_3$ increases, the ER increases while the SR decreases, indicating a trade-off between these two performance metrics. As $t_3$ increases, the Fano-asymmetric spectral lineshape gradually evolves into a horn-like lineshape with a reduced notch depth. The IL shows a slight decrease as $t_i$ ($i = 1$--$3$) increases.

\section{Device fabrication and characterization}

Directional couplers are essential in PICs for splitting and combining optical waves. They consist of two closely spaced waveguides, with coupling strength controlled by the interaction length or separation gap. 
Based on coupled mode theory~\cite{chrostowski2015silicon, arianfard2023sagnac}, the coupling relies on phase matching  condition between the two fundamental eigenmodes of the coupled waveguides, typically referred to as the even and odd modes, or symmetric and anti-symmetric modes. Figure 3(a) illustrates the mode profiles of the even and odd modes in a directional coupler, which is formed by two parallel silicon wire waveguides. The waveguides are designed with a width of 500 nm, a height of 220 nm, and a gap size of 100 nm for the directional coupler. Directional coupler lengths in our proposed device layout design are determined based on the relationship between the coupling coefficient and the coupling length, as described in Refs. \cite{arianfard2023sagnac, chrostowski2015silicon}. To establish a clear link between theoretical calculations and device fabrication, we employ the following design-to-layout workflow. The target coupling coefficients ($t$, $\kappa$) are first obtained from the analytical scattering-matrix model that produces the desired horn-like and Fano-like spectral responses. The even and odd supermodes of the directional coupler are then simulated (Fig.~3(a)) to determine the crossover length $L_x$, which corresponds to the propagation distance over which optical power fully transfers from one waveguide to the other, and is defined as

\begin{equation}
    L_x = \frac{\lambda}{2(n_{\text{eff,even}} - n_{\text{eff,odd}})} ,
\end{equation}

where $\lambda$ (here, $\lambda = 1550~\text{nm}$) is the operating wavelength, and $n_{\text{eff,even}}$ and $n_{\text{eff,odd}}$ are the effective indices of the even and odd supermodes, respectively. For a target field coupling coefficient $\kappa$, which generates the desired horn-like and Fano-like spectral responses (obtained from the analytical model based on the scattering-matrix method), the required coupling length $L_c$ can be determined as

\begin{equation}
    L_c = \frac{2L_x}{\pi} \arcsin(\kappa) - L_b ,
\end{equation}

where $L_b$ is the effective additional coupling length introduced by the bending waveguides on both sides of the directional couplers. The values of $L_b$ were derived from the measured power split ratios of the fabricated devices and applied in the next round of designs. For the horn-like device, $t_1 = t_3 = 0.89$, $t_2 = 0.99$, yielding $L_{c1} = L_{c3} \approx 8.3~\mu\text{m}$ and $L_{c2} \approx 2.41~\mu\text{m}$. For the Fano-like device, $t_1 = t_3 = 0.88$, $t_2 = 0.98$, giving $L_{c1} = L_{c3} \approx 8.67~\mu\text{m}$ and $L_{c2} \approx 3.45~\mu\text{m}$.

\begin{figure*}[ht]
    \centering
    \includegraphics[width=\textwidth]
    {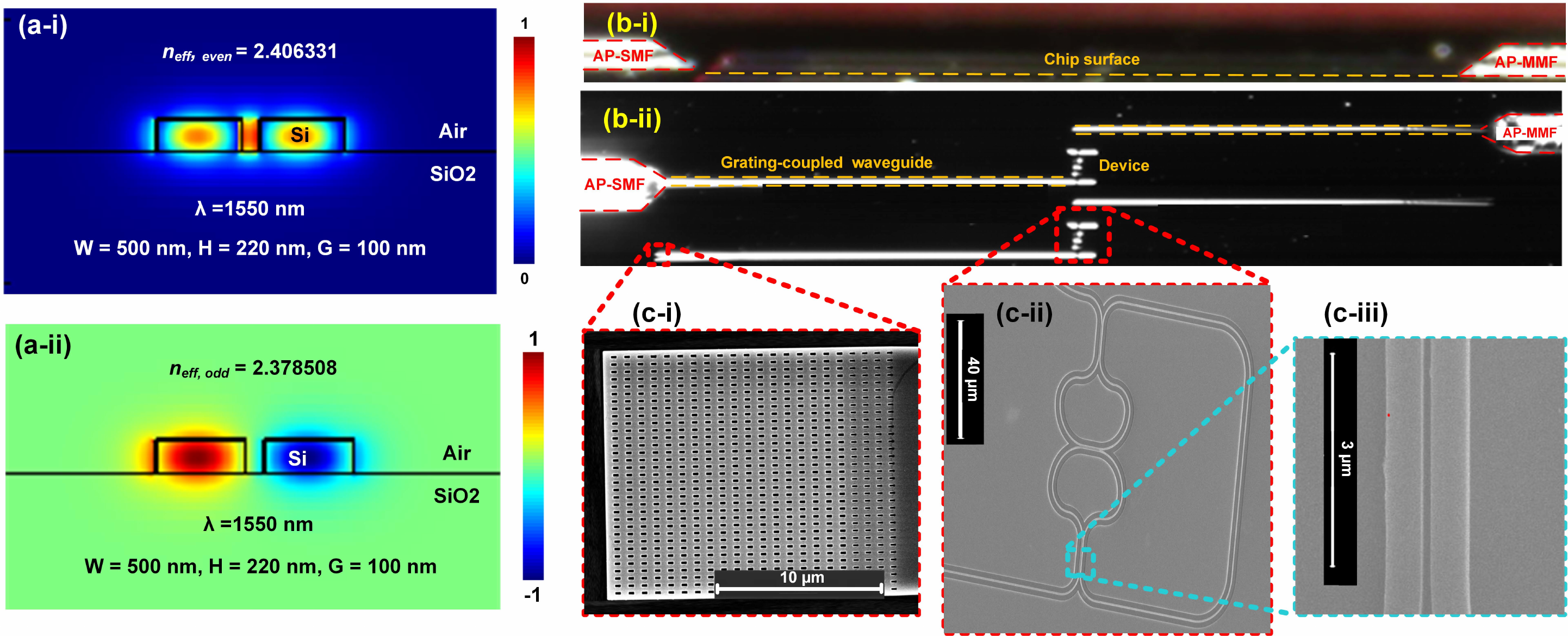} 
    \caption{(a) Designing directional couplers using two adjacent silicon wire waveguides. (i) and (ii) show the simulated mode profiles for the even and odd modes of the directional coupler, respectively. W: width. H: height. G: gap between the two waveguides in the directional coupler. (b) Macrograph of the coupling scheme utilizing angle-polished fibers and grating couplers fabricated on an SOI platform; (i) side view and (ii) top view of horizontally positioned fibers on the chip. AP-SMF: angle-polished single-mode fiber. AP-MMF: angle-polished multi-mode fiber. The devices shown in (b-ii) are replicated grating-coupled devices arranged in a column to improve yield and provide redundancy. (c) SEM images of the fabricated components: (i) the fully etched grating coupler consisting of two square arrays of rectangular holes, (ii) coupled SIs formed by a self-coupled wire waveguide, and (iii) a zoomed-in micrograph of the coupling region of the SI.}
    \label{fig:SEM}
\end{figure*}

The devices, designed based on the aforementioned principle, was fabricated on an SOI wafer featuring a 220 nm thick top silicon layer and a 2 µm thick buried oxide (BOX) layer. The device fabrication primarily follows standard complementary metal-oxide-semiconductor (CMOS) processes, with the exception of device patterning, which is achieved using electron-beam lithography. Macrograph of the two fabricated devices is shown in Figure 3(b-ii), while Figure 3(c) presents scanning electron microscopy (SEM) zoomed-in views of: (i) the grating coupler, (ii) the proposed coupled SIs formed by self-coupled nanowire waveguides, and (iii) the directional coupler.  During fabrication, electron-beam lithography (Vistec EBPG 5200) was employed to define the device layout on a positive photoresist (ZEP520A). This was followed by inductively coupled plasma (ICP) reactive ion etching (RIE) process to transfer the pattern to the top silicon layer, using SF$_6$ and C$_4$F$_8$ as etching gases. For efficient light coupling, two-step apodized grating couplers (Figure 3(c-i)), consisting of two square arrays of rectangular holes fully etched into the 220 nm silicon slab as described in \cite{benedikovic2014high}, were integrated at the ends of Port 1 and Port 2. These couplers facilitated the coupling of light into and out of the chip using angle-polished single-mode and multi-mode fibers, respectively, with the fibers positioned horizontally above the chip surface (Figure 3(b-i)). Various designs of grating couplers on the SOI platform have been demonstrated by both academic and industrial research groups \cite{cheng2020grating}. However, these designs typically rely on shallow etching of the silicon layer, which requires two lithography steps, thereby increasing fabrication complexity and cost. Fully-etched grating couplers present a simpler and more cost-effective alternative, as they enable the fabrication of photonic circuit components in a single lithography step, making them particularly suitable for rapid prototyping. Despite their advantages, fully-etched grating couplers face the significant challenge of strong back reflections into the waveguide. This challenge arises due to the high refractive index contrast in the grating region, which substantially increases Fresnel reflections compared to shallow-etched designs. To mitigate this issue, we employed the two-step apodized grating coupler detailed in \cite{benedikovic2014high}, which reduces back reflections.

\begin{figure*}[ht]
    \centering
    \includegraphics[width=\textwidth]{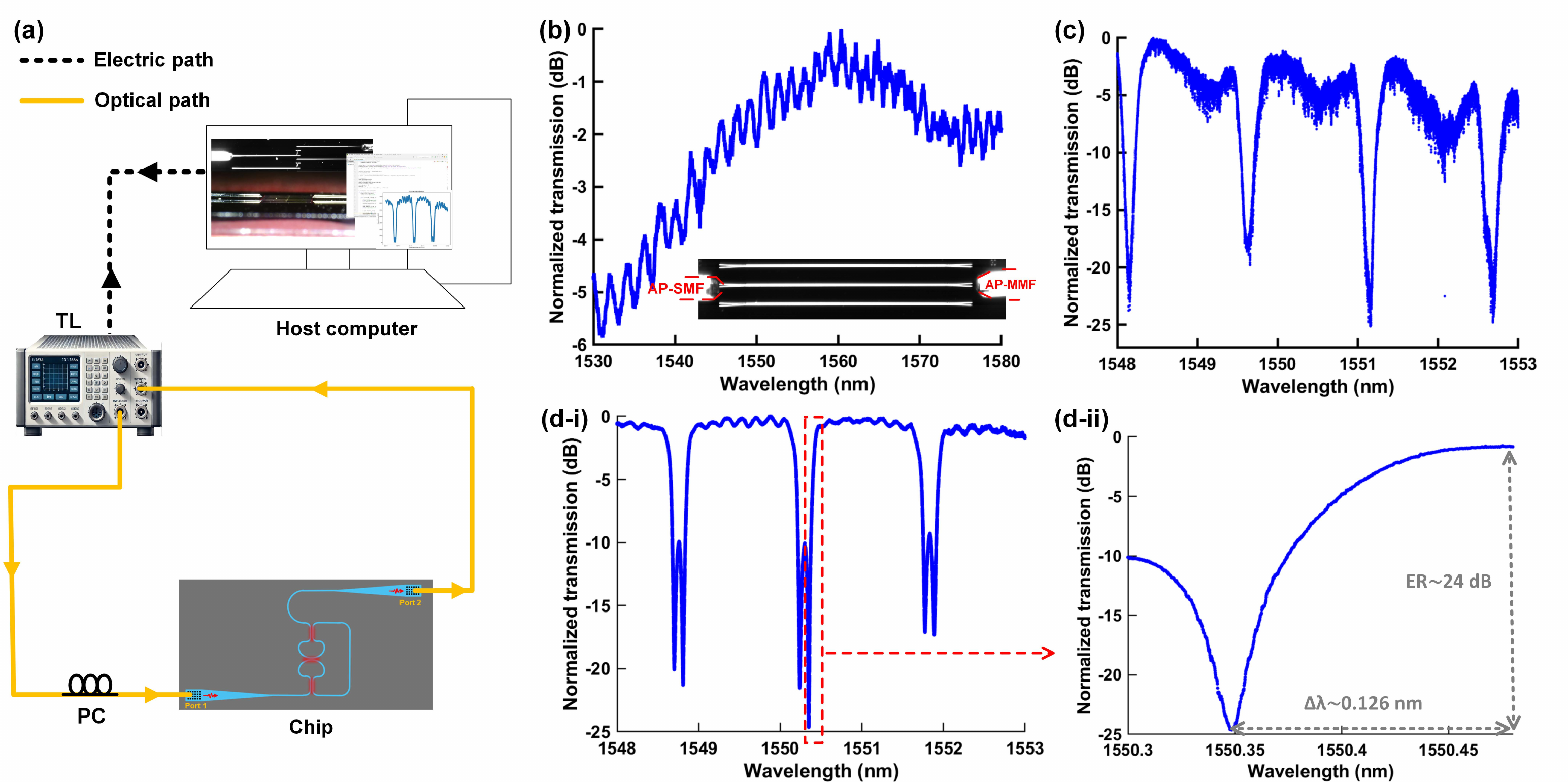} 
    \caption{(a) Experimental setup for transmission spectrum measurement using the tunable laser scanning method. TL: tunable laser. PC: polarization controller. (b) Transmission spectrum of a straight waveguide with a pair of grating couplers at both ends. The inset shows a macrograph of the top view of the coupling scheme for input and output of a straight waveguide, utilizing grating couplers and angle-polished fibers at both ends. AP-SMF: angle-polished single-mode fiber. AP-MMF: angle-polished multi-mode fiber. (c) Measured transmission spectrum of the device, fabricated based on the structural parameters for the horn-like spectral lineshape shown in Figure 2(c). (d-i) Measured transmission spectrum of the device, fabricated based on the structural parameters designed to achieve the Fano-like spectral lineshape shown in Figure 1(c). (d-ii) Zoomed-in view of (d-i) within the wavelength range of 1550.3 nm–1550.5 nm. Panels (c) and (d) show data measured from the devices using grating couplers and angle-polished fibers at both ends.}

    \label{fig:Measurement}
\end{figure*}

\begin{table*}[t]
\begingroup

\centering
\caption{Performance comparison of reported works.}
\label{tab:perf-compare}
\renewcommand\arraystretch{1.2}
\begin{tabular}{p{5.0cm} c c c c c}
\hline
\textbf{Scheme} & \textbf{Platform} & \textbf{ER (dB)} & \textbf{SR (dB/nm)} & \textbf{Footprint ($\mu$m$^2$)} & \textbf{Ref.} \\
\hline
Two MRRs embedded in a feedback loop & SOI & 29.2 & 83.42 & $\sim$604$\times$20 & \cite{zhao2019independently} \\
MZI with dual MRRs & SOI & 16.5 & $91$ & $\sim$435$\times$88 & \cite{guo2019twin} \\
Two MRR-coupled MZI & SOI & 56.8 & 3388.1 & $\sim$830$\times$340 & \cite{liu2019tunable} \\
Embedded MRRs & SOI & 40 & 93 & $\sim$35.8$\times$10 & \cite{wang2017slope} \\
Add-drop MMR-coupled FP cavity & SOI & 23.22 & 252 & -- & \cite{zheng2017compact} \\
Embedded MRRs & SOI & 48 & 102 & $\sim$120$\times$45 & \cite{wang2016fano} \\
MRR with feedback-coupled waveguide & SOI & 30.8 & 226.5 & $\sim$325$\times$20 & \cite{zhao2016tunable} \\
A grating-based FP cavity-coupled MRR & SOI & 22.54 & 250.4 & $\sim$132.7$\times$17.5 & \cite{zhang2016optically} \\
PhC cavity side-coupled to a waveguide FP & SOI & 7.3 & 5 & $\sim$6$\times$1 & \cite{yu2013fano} \\
All-pass MRR with an air-hole in the bus waveguide & SOI & 20 & 400 & $\sim$60$\times$60 & \cite{gu2020fano} \\
Two MMRs embedded in MZI & SiNOI & 57 & $8.1\times10^{4}$ & -- & \cite{cheng2022achieving} \\
MRR-coupled FP cavity & SOI & 13.7 & 510 & $\sim$705$\times$680 & \cite{zhang2022optical} \\
Add-drop MMR embedded FP cavity & SOI & 40 & 700 & -- & \cite{li2017actively} \\
Add-drop MDR embedded in MZI & SOI & 30.2 & 41 & $\sim$850$\times$390 & \cite{zhang2018thermally} \\
Two MMRs embedded in MZI & SOI & 6.44 & 247.54 & $\sim$454.84$\times$20.94 & \cite{liao2022chip} \\
Nonlinear interference within a waveguide-coupled MRR & GeSbS (chalcogenide glass) & 35.3 & $2.5\times10^{5}$ & $\sim$903$\times$905 & \cite{xiao2024tunable} \\
Add-drop MDR embedded in MZI & SOI & 17.1 & 13.6 & $\sim$432$\times$243 & \cite{zhou2022fully} \\
\makecell[l]{This work} & SOI &
\makecell[l]{24 (experiment)\\ 63 (theory)} &
\makecell[l]{190 (experiment)} &
$\sim$93$\times$101 & -- \\
\hline
\end{tabular}

\vspace{3pt}
\raggedright
\footnotesize
MRR: microring resonator. MZI: Mach--Zehnder interferometer. FP: Fabry--Perot. MDR: microdisk resonator.\\
SiNOI: silicon-nitride-on-insulator.

\endgroup
\end{table*}

To characterize the transmission spectra of the fabricated devices, the experimental setup shown in Figure 4(a) is used. These spectra were obtained by stepped-wavelength sweeping of a continuous-wave laser (Keysight/Agilent HP~8164A) across the desired wavelength range with increments of 0.1\,pm at an output power of approximately 1\,mW. The output powers from the devices under test were recorded using the laser’s built-in power sensor (HP~81532A). After each individual tuning step, the measured output power was extracted with an averaging time of 20\,ms. Figure 4(b) shows the transmission spectrum of a straight waveguide with 2D fully etched grating couplers at both ends. The ripples with a maximum ER of $\sim 1.7 \, \text{dB}$ in the spectral response arise from FP interference caused by back reflections between the grating couplers. The inset in Figure 4(b) presents a macrograph of the top view of the coupling scheme, illustrating the input and output of a straight waveguide. This setup utilizes grating couplers and angle-polished fibers positioned at both ends for efficient light coupling.
Figure 4(c) illustrates the measured transmission spectrum of the device, fabricated using structural parameters designed to achieve the horn-like spectral lineshape depicted in Figure 2(c). Figure 4(d-i) presents the measured transmission spectrum of the fabricated device, which was designed based on structural parameters intended to achieve the Fano-like spectral lineshape shown in Figure 1(c). A zoomed-in view of the spectrum within the wavelength range of 1550.3 nm–1550.5 nm, highlighting finer details of the spectral response with an ER of \(\sim 24 \, \text{dB}\) and a corresponding SR of \(\sim 190 \, \text{dB/nm}\), is shown in Figure 4(d-ii). The observed ripples in the passband of the measured spectra in Figures 4(c) and 4(d), compared to the calculated spectra, are primarily attributed to back reflections from the grating couplers at Port 1 and Port 2 of the device, which result in a steeper filtering roll-off. The horn-like device in Figure~4(c) shows better agreement with the simulation results (ER~$<$~30~dB) in Figure~2(c), while the Fano-like device (designed for ER~$\approx$~63~dB) exhibits a smaller measured ER and an apparent flattening of the resonance profile in the recorded transmission spectra (see the top of the spectrum in Fig.~4(c-i)), where the lower portion of the resonance dip is truncated. This behavior appears to result from a combination of back reflections from the grating couplers and the limitations of the laser-scanning measurement setup, which is not sufficient for accurately characterizing Fano resonances with ultra-high ERs. To overcome such issues, previous studies have employed a vector network analyzer-based experimental setup, which offers higher resolution and dynamic range, enabling more precise characterization of Fano resonances~\cite{liu2019tunable, cheng2022achieving}. In addition, minor spectral shifts and other discrepancies between simulation and experiment are attributed to fabrication-induced variations, such as etch-depth and coupling-gap deviations, which alter the effective coupling coefficients and phase balance within the device.
Table II compares representative experimentally demonstrated Fano-resonance structures, highlighting differences in configuration, platform, extinction ratio, slope rate, and footprint. Hybrid designs such as FP–coupled MMRs and MZI–assisted structures typically require precise wavelength alignment and occupy larger footprints, with performance sensitive to thermo-optic drift. In contrast, the proposed self-coupled waveguide–based architecture achieves a compact, high-performance design with relaxed alignment requirements and enhanced fabrication tolerance.

\section{Conclusion}

We demonstrated, both theoretically and experimentally, that a compact integrated photonic resonator, which employs only two SIs and a connected feedback waveguide, can generate high-performance periodic Fano-like resonances. This high performance is attributed to strong coherent optical mode interference within the compact resonator, which incorporates both IIR and FIR filter elements. These results highlight the capability of compact coupled SIs to generate high-performance Fano-like resonances, making them suitable for various applications in classical and quantum photonics.

\section*{Acknowledgements}
AP acknowledges an RMIT University Vice-Chancellor’s Senior Research Fellowship and a Google Faculty Research Award. This work was supported by the Australian Government through the Australian Research Council under the Centre of Excellence scheme (No: CE170100012).
The authors acknowledge the facilities and the scientific and technical assistance provided by RMIT University’s Microscopy \& Microanalysis Facility, a linked laboratory of Microscopy Australia enabled by NCRIS, as well as the Micro Nano Research Facility (MNRF). This work was performed in part at the Melbourne Centre for Nanofabrication (MCN) in the Victorian Node of the Australian National Fabrication Facility (ANFF). This work was supported by RMIT University's SoE Support Crazy Idea E\&E (ID: PROJECT\_PLAN\_TASK-3-72725).

\section*{Author Contributions}
Hamed Arianfard designed, fabricated, and characterized the device and prepared the manuscript. Tim Weiss, Yang Yang, and Joshua Bader contributed to the characterization. Alberto Peruzzo supervised the project. All authors participated in the review and discussion of the manuscript.
 
\section*{Disclosures}
The authors declare that there are no conflicts of interest.

\bibliographystyle{unsrt}
\bibliography{references}

\end{document}